# Electron matter waves with internal torque

Y. Fang, J. Kuttruff, P. Baum*

*Universität Konstanz, Fachbereich Physik, 78464 Konstanz, Germany*
*peter.baum@uni-konstanz.de*

**Angular momentum and torque are important principles for basic and applied physics on any spatial scales, for example, in elementary particles [1], cold gases [2], optical tweezers [3], quantum information technology [4], metamaterials [5], gyroscopes [6] or astrophysical entities [7]. Investigating or controlling angular momentum in atoms or sub-atomic structures requires torque on femtosecond and picometer scales, far below the capabilities of laser light [8-12]. Here we shape the electrons in an electron microscope into wave packets with a time-dependent chirality and internal torque. We intersect the electron beam with chiral laser light to create discrete energy sidebands [13,14] by multiple helical photon absorptions that create a correlation between orbital angular momentum and kinetic energy. Dispersion of these partial waves due to the electron's rest mass [15] then converts each single electron into a wave function with internal torque. Under our control, a left-handed matter wave becomes right-handed in femtosecond times. Such quantum objects will facilitate research on angular momentum and chirality on atomic and sub-atomic scales.**

Investigating the structure and dynamics of materials on atomic dimensions requires access to the motion of atoms and electrons in space and time [16,17]. If rotation is involved, torque must be applied [18]. Shaped laser pulses with chiral electromagnetic field can exert angular momentum onto a material via the photon spin [8], phase-vortex beams [9] or, in the most general case, optical beams with self-torque [10-12]. However, coupling the photon torque with an individual atom inside a material is challenging because the wavelength of laser light lies in the nanometer range, far above the size of atoms or molecules. Even if higher harmonics are applied [10], their wavelength is still too long to probe or control the rotation of a material on the level of the single atoms or unit cells. However, in contrast to the photons of laser light, electrons carry a rest mass and can therefore easily have a wavelength in the picometer range, ten times smaller than an atomic size. Structured electrons in which the wave function is shaped in space and time [19,20] are therefore useful objects for studying and manipulating materials on atomic scales [21-24]. In terms of angular momentum, researchers have so far demonstrated electron vortex beams [25-30], twisted currents [31,32] and coils of charge and mass [33], but a matter wave with a time-dependent chirality remains to be observed, although it would be a useful and general concept to investigate and control the ultrafast dynamics of angular momentum on an atomic scale.

Here, we report the creation and observation of electrons with an ultrafast internal torque. Figure 1 depicts our concept and experiment. We use an ultrafast transmission electron microscope with a Schottky field emitter source [34,35] at an electron energy of $E_0 = 70$ keV. Femtosecond laser pulses (green) hit the emitter tip (grey) and create a femtosecond electron beam. We use less than 0.1 electrons per pulse to avoid temporal space charge effects [35]. The formation and measurement of self-torqued electrons are realized by three steps (see Fig. 1a). We first devise a chiral coherent quantum walk that creates correlated energy-vortex states (light red area). We then let the wave packet disperse in free space (light blue area) and finally characterize the resulting matter wave in space and time.

In our chiral quantum walk, we intersect the electron beam with a laser beam at an ultrathin membrane as a modulation element. The resulting coherent laser-electron interaction [13] imprints the matter wave of the electrons with a series of coherent sidebands in the energy domain [13,14]. However, we aim here for energy sidebands with orbital angular momentum of integer amount. We therefore apply a first-order, left-handed optical vortex beam (red) that carries an optical orbital angular momentum of $\ell_{ph}\hbar$ per photon, where $\ell_{ph} = 1$. The laser wavelength is $\lambda_{ph} = 1030$ nm, the frequency is $\omega = 2\pi c/\lambda_{ph}$ and the photon energy is $E_{ph} = \hbar\omega \approx 1.2$ eV, where $c$ is the speed of light. The free-standing silicon nitride membrane (brown) changes the optical phase of the laser beam upon transmission and therefore imprints an instantaneous momentum change onto the electron de Broglie wave [36]. If the temporal coherence of the electrons exceeds the optical cycle period [37], spectral interference creates a series of discrete and coherent electron energy sidebands at integers of the photon energy [13,14]. In our chiral photon-electron interaction, angular momentum must be conserved, and the absorption of one chiral laser photon implies the creation of one quantum of orbital angular momentum in the electron's matter wave. Multiple photon-absorption and photon-emission pathways are known to create a complex shape of intensities [14], but the correlation of final energy with final angular momentum should always be maintained. The $n$-th electron partial wave at an energy of $E_0 + n\hbar\omega$ has in sum absorbed $n$ photons from the optical vortex beam and should therefore carry an orbital angular momentum of $n\hbar$ per electron (see Fig. 1b). A negative $n$ corresponds to an effective emission of chiral photons into the optical beam. After the quantum walk, we therefore expect that each sideband acquires a left-handed or right-handed helical phase of its matter wave (blue spirals in Fig. 1a) in proportion to the number of absorbed or emitted laser photons. Due to the singularity in the middle, the electron beam profile should acquire a donut mode [25,26] of increasing clarity (blue rings in Fig. 1a).

Next, the resulting electron wave function acquires dispersion by propagation in free space. Higher-energetic parts travel faster in vacuum than lower-energetic parts [15]. For small energy changes $\Delta E = n\hbar\omega$ around $E_0$, the temporal separation is $\Delta t(n) = -nLE_{ph}m_{el}^{1/2}[E_0(\gamma+1)]^{-3/2}$, where $L$ is the propagation distance, $m_{el}$ is the rest mass of the electron, and $\gamma = 1.137$ is the Lorentz factor. Figure 1c shows the calculated arrival times of the different orbital angular momentum parts after $L = 18$ cm of propagation in free space. We can see a well-discernible but also overlapping temporal separation and therefore a continuous coupling between orbital angular momentum and time. A partial wave with a positive orbital angular momentum (for example $3\hbar$) arrives earlier than a partial wave with a smaller or negative amount (for example $-3\hbar$). Hence, we expect the creation of an overall electron matter wave with a changing angular momentum and internal torque.

In the experiment, we first verify the formation of a discrete amount of orbital angular momentum as a function of the electron energy. Figure 2a shows the measured spatial profile of the modulated electron beam without any energy filtering. We see a Gaussian-like distribution with a full-width-at-half-maximum of ~4 μm. The energy spectrum (Fig. 2b) shows coherent energy sidebands from −5th to 5th order in agreement between measurement (black dots) and theory (blue line). The Bessel functions of a plane-wave interaction [14] are smeared out because the donut-shaped intensity profile of our optical vortex beam produces a varying electron-photon coupling strength across the beam. However, each final energy still corresponds to a well-defined total number of chiral photons that are emitted or absorbed. Figure 2c shows the simulated helical phases of the partial de Broglie waves. The singularity in the middle must correspond to zero intensity.

Figure 2d shows the measured spatial profiles of the individual electron partial waves at energies of $E_0 + n\hbar\omega$, obtained with an imaging electron energy analyzer (see Methods). The 0th electron partial wave at $E_0$ is concentrated in a small spot with a width of ~0.5 μm. The higher-order partial waves reveal donut-like profiles with increasing ring diameters, indicating post-selected partial electron beams with an increasing amount of orbital angular momentum, similar to beams from forked diffraction gratings [27], magnetic monopoles [28] or circular plasmon waves [29]. Slight elongations are caused by the electron-membrane angle of 56° that projects the round optical vortex field onto the round electron beam with an asymmetric aspect ratio. Rotated axes come from the microscope's magnetic lenses. Figure 2e shows a wave-optical simulation that includes these spatial propagation effects (see Methods). The agreement between theory and experiment provides two central insights. First, the electron partial wave in the $n$th sideband carries an orbital angular momentum $n\hbar$ per electron. Second, the electron pulses cover the entire optical vortex beam in a coherent way. Only some slight halos in the measured partial waves indicate a minor role of incoherent electrons.

To determine how our different electron sidebands are dispersed in time, we measure the relative arrival time at a distance of 18 cm behind the generation stage with a second laser-induced quantum walk. Figure 3a shows the configuration of this measurement. A plane-wave probe laser (red) intersects with the shaped electron beam at a second silicon nitride membrane that mediates the optical interaction with the electrons, but no angular momentum is applied, and the interaction is made weak (see Fig. 3b). Scanning the arrival time of the probe laser pulses with respect to the shaped electrons then reveals for each sideband $n$ its arrival time. Figure 3c illustrates the details of our secondary quantum walk. Electrons are transferred from each sideband $n$ to the two nearby sidebands $n \pm 1$ [38,39]. Already existing energy sidebands obtain or loose intensity in multiple ways with a complex dependency on time, but new higher-order sidebands outside the initial spectrum [38] can only appear at such time delays at which the original highest-order sideband of the self-torqued beam has arrived. The timing of lower-order sidebands in the modulated beam can be characterized by reduction of the laser power in the generation stage. We expect dispersion on femtosecond scales, and therefore we avoid attosecond interferences [40] by tilting the silicon nitride membrane in the generation stage away from velocity match (see Methods).

Figure 3d shows the measured electron spectrum as a function of $\Delta t$; see Extended Data Fig. 1 for additional results. At $|\Delta t| > 300$ fs, the characterizing laser beam does not interact with any of the chiral sidebands, and we see the original spectrum from the generation stage (compare Fig. 2b). Around $\Delta t = 0$, the incoming chiral sideband spectrum gets additional broadening that emerges for different initial sidebands with distinct time delays (blue dots). The slight chirp that our initial femtosecond electron pulses have obtained from acceleration in the electron source [34] is not relevant for this metrology (right column, Extended Data Fig. 1) because it only tilts the sidebands themselves, not the distribution of arrival times. Figures 3e and 3f show the details of the measured arrival time data for final $n = \pm 3$ and $n = \pm 4$. A peak shift of 30 fs for the $\pm$3rd energy sidebands (Fig. 3e) implies a 30-fs delay between the +2nd and the −2nd partial waves of the incoming shaped electron. The delay between the −3rd and the 3rd partial waves is 45 fs (Fig. 3f). Figure 3g shows a summary of all time-domain results. We see a linear delay of the sideband arrival time with respect to $n$ that matches well with analytical dispersion theory (dashed line). Due to the proportionality between $n\hbar\omega$ and $n\hbar$ (see Fig. 2), angular momentum also arrives linearly with time.

Therefore, the combined results show that the electrons in our experiment are indeed converted into a matter wave with internal torque (see Fig. 4a). The local chirality within our electrons changes from left-handed at the beginning into achiral in the middle towards right-handed in the end. Figure 4b and 4c show the orbital angular momentum and the internal torque as a function of arrival time of the electron wave at a material (see Methods). The observed time-dependencies and absolute magnitudes (blue) agree with simulation results (black). The orbital angular momentum changes from $-5\hbar$ to $5\hbar$ within 400 fs at a peak torque of ~65 $\hbar$/ps. In the spatial domain, this electron self-torque has a ring shape (Extended Data Fig. 2). Although the eigenstates of orbital angular momentum are integer multiples of $\hbar$, the relative weights of their superposition (see Fig. 1c) make the instantaneous angular momentum continuous, and we obtain a smooth gradient in time and space.

Our concept and experiment provide several ways to control the amount, range and timing of the angular momentum and thereby the properties of the internal torque. For example, if we adjust the intensity of the optical vortex beam or its phase order $\ell_{ph}$, we can control the bandwidth of the generated sideband spectrum and thereby the range of angular momentum in the beam. If we apply an intermediate energy filter after beam formation, or control our electrons with multi-color laser fields [38], we can control the internal shape of the sideband spectrum to customize the time-dependence of the resulting self-torque wave. If we alter the central energy $E_0$ of the electron beam or the propagation distance $L$, we can control how the shaped energy spectrum converts into an internal torque in time. If we adjust the coherence length of the electrons [37], we can tune between a smooth gradient (Fig. 4b-c) and a temporally localized torque that changes step by step (Extended Data Fig. 3). If we generate self-torque electrons with a continuous laser wave [41], every single electron will still obtain internal torque. Our electrons can therefore be produced at the full beam brightness of modern electron microscopes.

In contrast to laser beams and photons with related properties [10-12], our electrons have a substantially smaller wavelength below atomic diameters. Also, they carry a rest mass and an electric charge. Their interaction with chiral materials is therefore complementary to optical or x-ray experiments. If our self-torque electrons are absorbed or scattered by an atom or molecule, we expect that they will exert a positive and then a negative torque onto an electronic state or a nuclear core in a very short time, inducing a finite impulsive rotational displacement that can serve as a trigger or probe of the subsequent evolution of the material. Post-selection of electron energy loss can reveal if a valence electron, an inner electron or a nuclear core has been modified or probed. This kind of electron-optical rotational tweezer on atomic scales should be useful, for example, to understand the mechanisms behind ring currents [42,43] or charge migration [44] in atoms, molecules or materials. Specially designed magnetic lenses can convert our beam into an ultrafast angular momentum streak in space [45] that might be useful for manipulating left and right parts of a nanostructure or molecule at different times. It will also be interesting to see how a collision between differently shaped electrons transfers angular momentum through twisted Coulomb effects. Sidebands with integer values of angular momentum under control of laser light may serve as sensors or carriers of information in free-electron quantum technology [39,46]. The zero-loss peak after modulation with optical vortex light is eight times smaller than the original electron beam, a feature that is potentially useful for ultrafast electron beam shaping attempts [29,47]. If we make the laser-electron interaction much more efficient, for example by whispering gallery modes [48], we might become capable of seeing the imprint of the electron beam's self-torque onto the photons of the laser beam. In general, the ability to shape an electrically charged and mass-bearing elementary particle like the electron into an almost arbitrary rotational form will enable to transfer almost any chiral-optical or dichroic principle of laser spectroscopy to the domain of electron microscopy, for studying chirality with atomic resolution in space and time.

**Acknowledgements:** We acknowledge funding from the Humboldt foundation and from the Deutsche Forschungsgemeinschaft (DFG) through SFB1432.

**Author contributions:** All authors designed the experiment, measured and analyzed data, and wrote the manuscript.

**Data availability:** The data supporting the findings of this study are available from the corresponding author upon request.

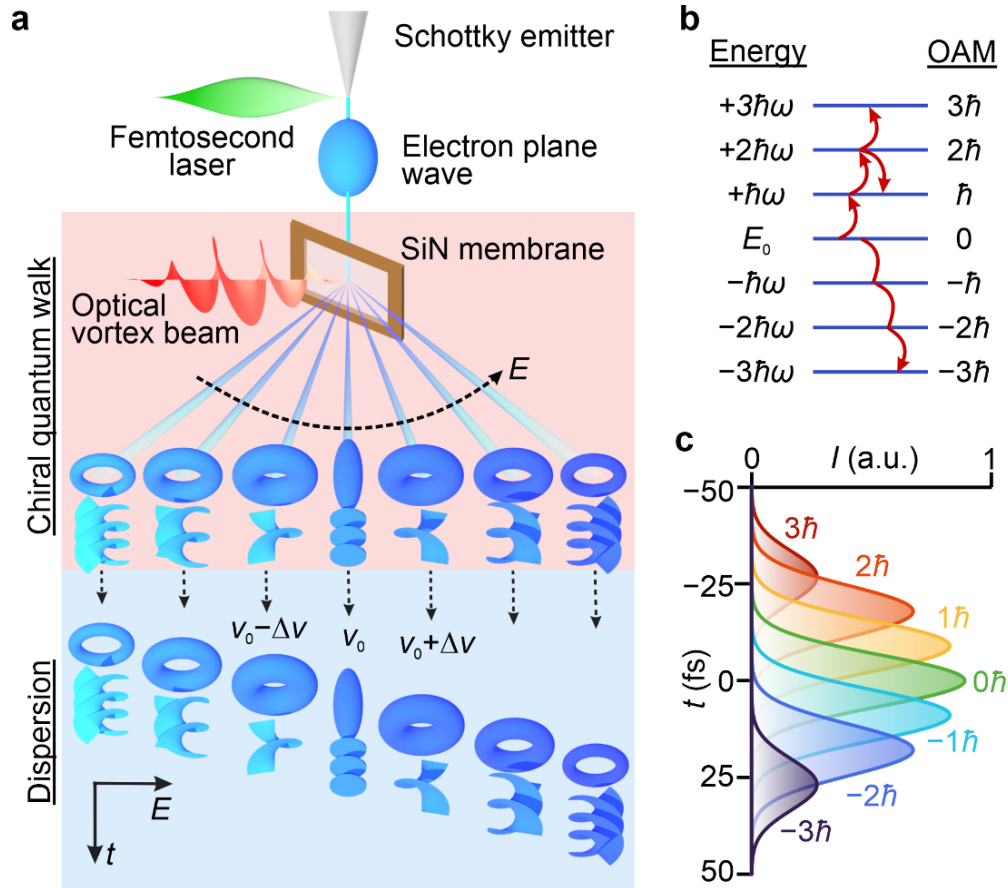

**a**, Experimental scenario. Ultrashort electron pulses (blue) are generated by photoemission with an ultrashort laser beam (green). Next, we apply a chiral quantum walk (light red area) to modulate the electron matter wave (blue) by an optical vortex beam (red). A SiN membrane (brown) provides the necessary interaction for discrete energy gain or loss. The blue rings indicate the resulting charge density and the spiral profiles indicate the phase of the de Broglie wave. Angular momentum becomes a function of electron energy $E$. Then, we apply a dispersive propagation in free space (light blue area). The velocity difference $\Delta v$ of the different electron partial waves separates them in time. In this way, we form an electron pulse with internal torque. **b**, Chiral quantum walk. In the interaction of a chiral laser photon with a nonchiral electron, energy gain or loss becomes proportional to gain or loss of orbital angular momentum (OAM). The red arrows show several possible transition paths. **c**, Dispersion of the resulting electron partial waves after propagation. Electrons with different angular momentum (colors) are separated in the time domain.

**Fig. 1. Concept of generating an electron beam with internal torque.**

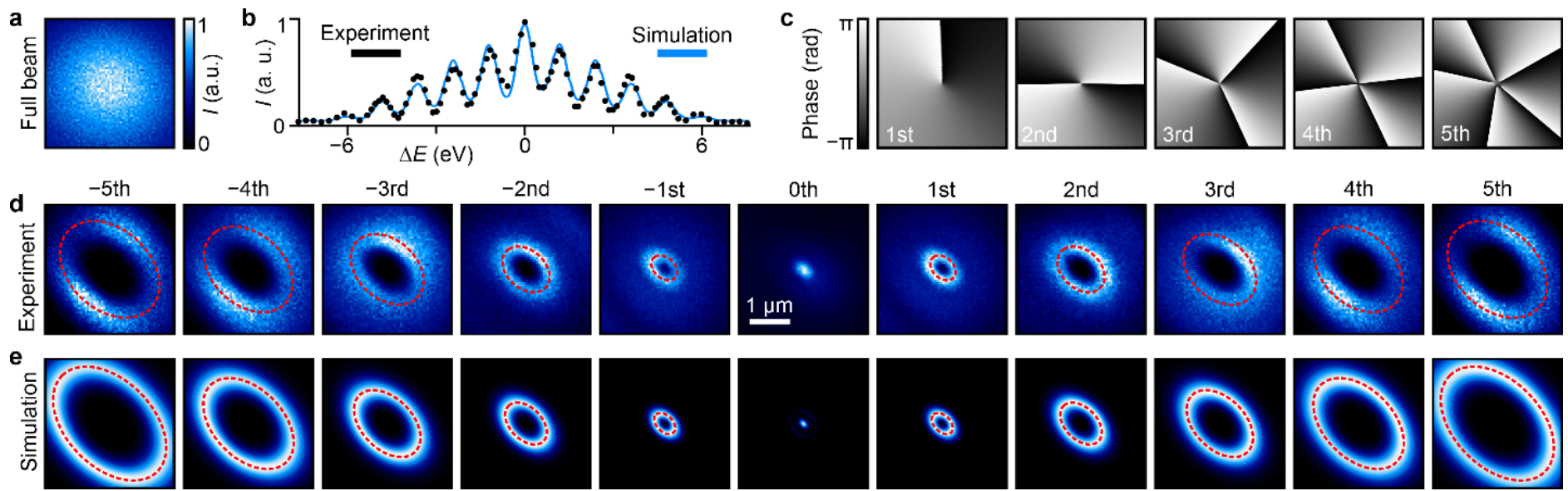

**Fig. 2. Energy spectrum and orbital angular momentum of electron partial waves. a**, Measured electron beam without filtering its energy. **b**, Measured electron energy spectrum (black dots) in comparison to a simulation (blue line). **c**, Expected helical phase of the de Broglie wave for different electron energies. **d**, Measured energy-filtered electron microscopy images of the individual electron partial waves. Red dashed lines, ring shapes. Scale bar, 1 μm. **e**, Simulation results.

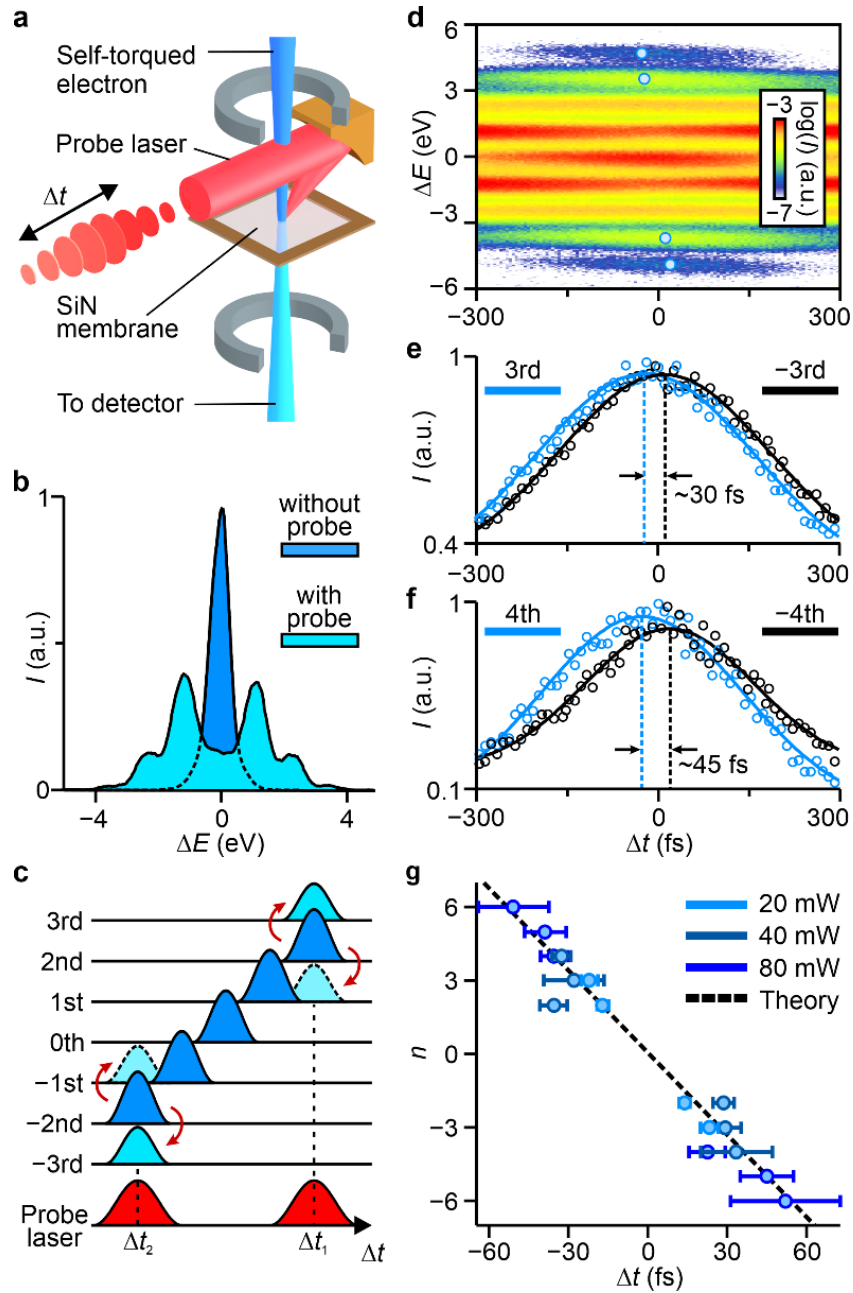

**Fig. 3. Time delay between the partial waves. a**, Characterization scheme. A weak plane-wave probe laser beam (red) overlaps with the incoming self-torque electron beam (blue) at a SiN membrane (brown). By scanning the arrival time $\Delta t$ of the probe laser, we measure the time of arrival of the partial waves. **b**, At temporal overlap, the spectrum without probe laser (dark blue) obtains neighboring sidebands and additional energies (light blue). **c**, Measurement principle. Different angular momentum sidebands (dark blue) arrive at different times. With the probe laser (red), we redistribute each sideband into neighboring ones (red arrows). Lower-order sidebands (dashed lines) can be reached in many ways, but higher-order sidebands (solid lines) can only be created if the probe laser arrives at the proper time (for example, $\Delta t_1$ for transfer from 2nd to 3rd sideband, or $\Delta t_2$ for transfer from -2nd to −3rd sideband). **d**, Measured electron spectrum as a function of $\Delta t$ for a laser power of 20 mW. Blue points indicate peak arrival times. See Extended Data Fig. 1 for additional results. **e**, Integrated intensity of the ±3rd sideband as a function of $\Delta t$. Circles, measurement; solid curves, Gaussian fit; dashed lines, fitted peak arrival times. The delay between the ±2nd electron partial waves is ~30 fs. **f**, Integrated intensity of the ±4th sideband and difference in arrival time. The delay between the ±3rd electron partial waves is ~45 fs. **g**, Compilation of all time-delay results (circles) in comparison with analytical theory (dashed line) as a function of the applied laser beam energy (colors).

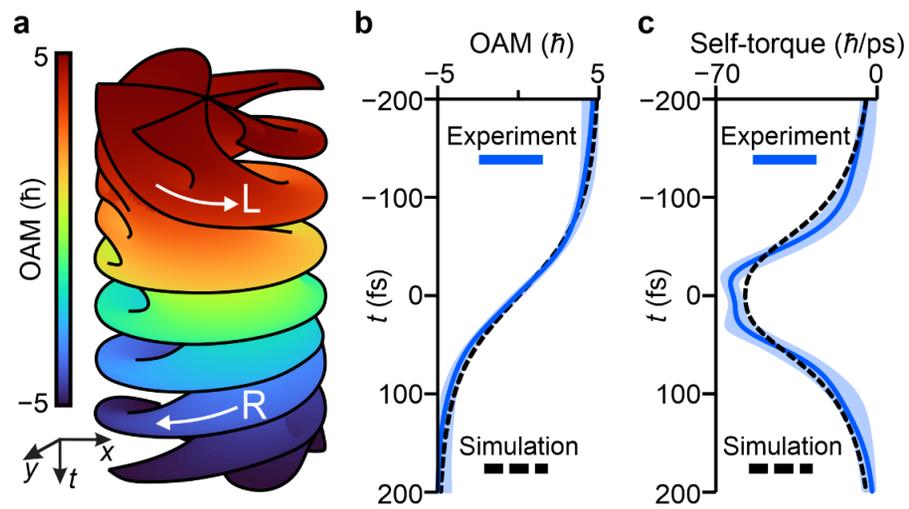

**a**

OAM (ℏ)

**b**

OAM (ℏ)

Experiment

Simulation

**c**

Self-torque (ℏ/ps)

Experiment

Simulation

**Fig. 4. Electron matter waves with internal torque. a**, Schematic of the phase of our electrons, showing a left-handed chirality (L) at the beginning of the pulse and a right-handed chirality (R) later on. The instantaneous orbital angular momentum is depicted with the color scale. **b** Orbital angular momentum (blue line) as a function of arrival time, in comparison to theory (dashed line). **c**, Internal torque (blue solid line) as a function of $t$, in comparison to theory (dashed line). We assume a 20 fs coherence time (compare Extended Data Fig. 3). The light blue areas show the statistical error of the experimental results.

## Methods

**Experiment:** We operate the experiments with an ultrafast transmission electron microscope with a Schottky field emitter source (JEM F200, JEOL) at an electron energy of 70 keV. The femtosecond laser (Carbide, LightConversion) has 1030 nm wavelength, 2 MHz repetition rate and the pulses are 250 fs long. The laser beam is split into three paths. In the first path, the beam is frequency-doubled with a beta-barium-borate crystal to trigger photoemission from the Schottky field emitter [35]. The electron pulse duration is ~270 fs [35] and the energy spread is about ~0.6 eV (full width at half maximum). We use less than 0.1 electrons per pulse. In the second path, the laser beam is converted into an optical vortex beam by a spiral phase plate with $\ell_p = 1$ (VortexPhotonics) and focused into the electron microscope with an $f = 350$ mm lens. The focal spot size is ~200 μm (diameter of $1/e^2$ intensity). The SiN membrane is 50 nm thick. The polarization of the optical vortex beam points along the direction of the electrons, and the angle between the electron beam and the membrane is ~56°. This angle deviates from the velocity-matching angle $\theta = \mathrm{atan}(c/v_{el}) = 25.4°$ [36], where $v_{el}$ is the velocity of the electrons, so that no two-stage attosecond interferences are produced [40]. In the third laser path, an off-axis parabolic mirror focuses the beam onto another 50-nm-thick silicon nitride membrane at a distance of $L = 18$ cm. The probe laser is approximately a plane wave with a diameter of ~8 μm (full width at half maximum), and the average power is ~1 mW. Finally, the electrons are guided into a magnetic post-column electron energy analyzer (CEFID, CEOS) and measured with an event-based direct electron detector (Timepix3, Amsterdam Scientific Instruments) that can measure every single electron [49]. The energy-filtered images of Fig. 2 are acquired with a 1-eV energy window in the energy plane of the spectrometer. In Fig. 3 and Extended Data Fig. 1, we scan the time delay of the probe laser in 5-fs steps, expose for ten seconds and repeat the scan five times for each data set.

**Data processing and fit:** In Fig. 3, we minimize the influences of laser and electron beam instabilities by normalizing the spectrum at each $\Delta t$ by its total intensity. Then, we use a nonlinear least squares algorithm to fit the raw data in the time domain with a Gaussian-like function, $y_n = a_1 \times \exp\{-[(\Delta t - a_2)/a_3]^2\} + a_4$, where $y_n$ is integrated over an energy interval of 1 eV to select the $n$-th-order sideband. The four fitting parameters, $a_1$, $a_2$, $a_3$, and $a_4$, correspond to the amplitude, peak moment, standard deviation, and intensity offset, respectively.

**Simulations of self-torque electrons in time:** We simulate the chiral quantum-walk of Fig. 1 by solving the relativistically corrected Schrödinger equation, $i\hbar\partial_t\psi(z,t) = \hat{H}\psi(z,t)$ [37]. The Hamiltonian is given by $\hat{H} = \hat{H}_0 + \hat{H}_1$, in which $\hat{H}_0 = E_0 + (\hat{p} - p_0)v_0 + (\hat{p} - p_0)^2/(2\gamma^3 m_{el})$ is the non-perturbed

Hamiltonian and $\hat{H}_1 \approx -eA_z\hat{p}/(\gamma m_{el})$ is the interaction part. Here, $\psi(z,t)$ is the electron wavefunction; $z'$ is a local coordinate around the center of the beam, $E_0$ is the central electron energy, $\hat{p}$ is the momentum operator, $p_0$ is the central electron momentum, $v_0$ is the central velocity of the electrons, $\gamma$ is the Lorentz factor, $m_{el}$ is the rest mass of electron, $e$ is the charge of the electron, and $A_z$ is the vector potential of the local laser field. After solving the quantum-walk interaction, we obtain a phase-modulated electron wavefunction $\psi(z')$. In the modulation, we account for multi-reflections between the front and back surfaces [36]. For propagation, we use a Fourier transform to obtain the electron wavefunction in momentum space, $\tilde{\psi}(k) = \int_{-\infty}^{\infty} \psi(z')\exp(-ikz')dz'$. Replacing $k$ with the electron energy, we obtain the spectrum of the electrons (see Fig. 2b). Then, we solve the Schrödinger equation for free propagation, $i\hbar\partial_t\psi(k) = \hat{H}_0\psi(k)$ to obtain the electron wavefunction $\psi_L(k)$ at a distance $L$. The electron wavefunction in temporal or longitudinal coordinates is calculated by applying the inverse Fourier transform, $\psi(z') = \int_{k_2}^{k_1} \psi(k)\exp(ikz')dk$. By selecting the upper limit $k_1$ and lower limit $k_2$ of the integration, we can inspect the temporal properties at selected energy sidebands.

**Simulation of vortex rings**: To account for the spatial shape of the optical vortex beam, we repeat the above calculation at each point in the focal plane of the laser beam. The optical field amplitude and phase are given by the Laguerre-Gauss modes of the laser beam and a $\sin^2$-shaped temporal envelope. We then simulate the spatial intensity of the electrons after free-space propagation by approximating the magnetic lens system in our electron microscope between generation and characterization stage as a single focusing lens. Utilizing Fraunhofer diffraction theory [50], the electron wavefunction at the focal plane, $\mathbf{r} = (x_f, y_f, z_f)$, is given by

$$\psi_f(x_f, y_f, z_f) = \frac{e^{i\alpha}}{i\lambda_{el}f}\iint \psi_0(x, y, z=0)\exp[-2\pi i(xq_x + yq_y)]dxdy \tag{1}$$

where $f$ is the focal length of this effective lens, $\lambda_{el}$ is the de-Broglie wavelength of electron, $\psi_0(x, y, z=0)$ is the wavefunction in the incident plane of the effective lens, obtained by solving the relativistically corrected Schrödinger equation at each point in focal plane, $q_x = x/(f\lambda_{el})$ and $q_y = y/(f\lambda_{el})$ are the components of the spatial frequency, the parameter $\alpha$ is $\alpha = [\pi h'_p(z)/\lambda_{el}h_p(z)](x_f^2 + y_f^2)$ in which $h_p(z)$ is a particular solution of paraxial equation of the electron trajectory in the vicinity of the optical axis. We optimize $f$ to fit the ring size of the simulation result with our measurement (see Fig. 2d-e).

**Reconstruction of orbital angular momentum and self-torque:** In Fig. 4, we extract the intensity of all electron partial waves from Fig. 2b. The timing of the partial waves is read from Fig. 3g, where the missing experimental data for ±1st and 0th partial waves are obtained by interpolation and theory (dashed line). This intensity and timing together with an estimated coherence time [37] are then used to reconstruct the orbital angular momentum and the internal torque.

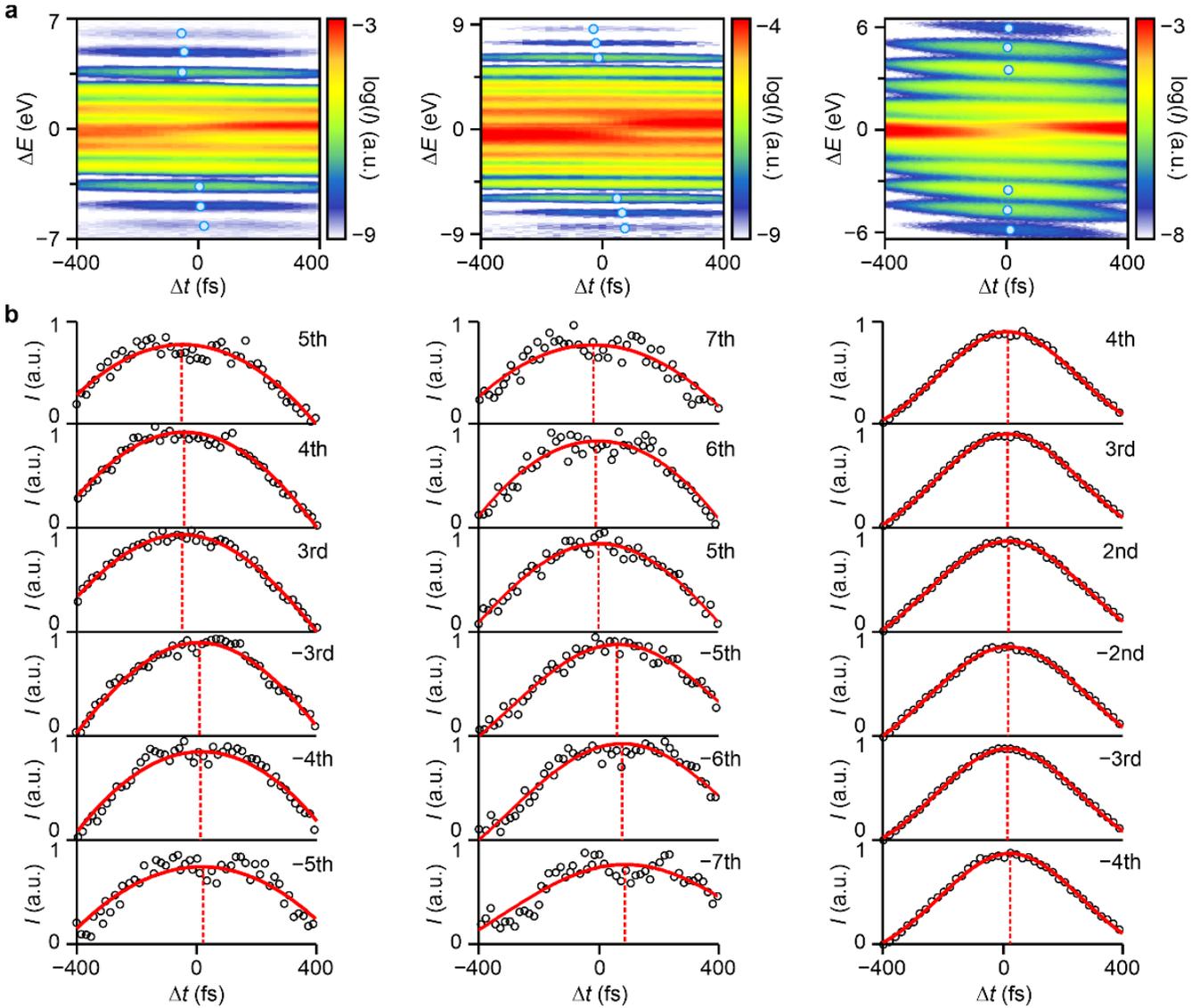

**Extended Data Fig. 1 | Additional time-delay measurement. a,** Time-delay measurement with a 40-mW optical vortex beam (left column), an 80-mw optical vortex beam (middle column) and no optical vortex beam (right column). Blue points indicate the measured peak arrival times. **b,** Details of the analysis. Electron intensity of different sidebands as a function of $\Delta t$. Circles, measurements; red solid curves, Gaussian fits; red dashed lines, indication of peak arrival time. The dips in higher-order sidebands stem from residual Rabi oscillation in the characterizing quantum walk [38]. Without optical vortex beam (right column), the peak arrival times for all sidebands are almost the same, indicating that the intrinsic chirp of the incoming electron pulses from the source (tilted individual sideband shapes) has negligible effects for our time-delay measurements.

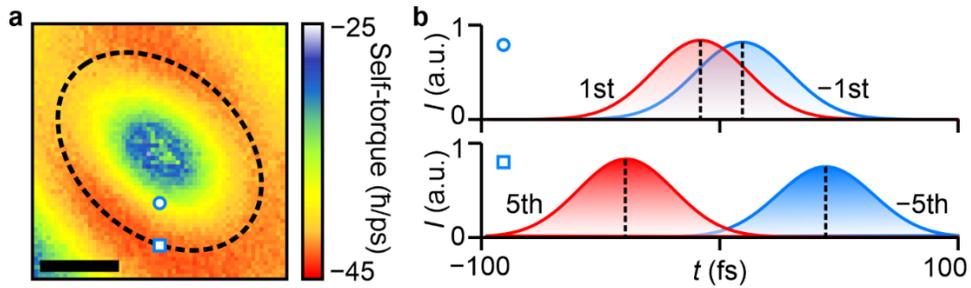

**Extended Data Fig. 2 | Two-dimensional distribution of the internal torque. a**, Spatially resolved self-torque extracted from the experimental results. Self-torque varies by approximately a factor of two. For each spatial position, the measured self-torque is averaged from $t = -100$ fs to $t = 100$ fs. The black dashed ellipse shows the donut-like profile of ±5th electron partial wave (compare Fig. 2d). The ring size of the self-torque is comparable to that of the ±5th electron partial wave. Squared and circle points indicate selected positions for further analysis. Scale bar, 1 μm. **b**, Interpretation of the measured self-torque shape. As shown in Fig. 2d, the electron partial waves are separated in space because of different ring sizes. At a moderate radius (white circle in a), the ±1st electron partial wave dominates, whereas at a larger radius (the white square point in a), the ±5th electron partial wave plays a more substantial role. For the ±1st sideband waves, the temporal profiles partially overlap in time and thus the overall orbital angular momentum is partially cancelled out. In contrast, the ±5th waves almost do not overlap in time, and the local self-torque is more prominent (compare Extended Data Fig. 3).

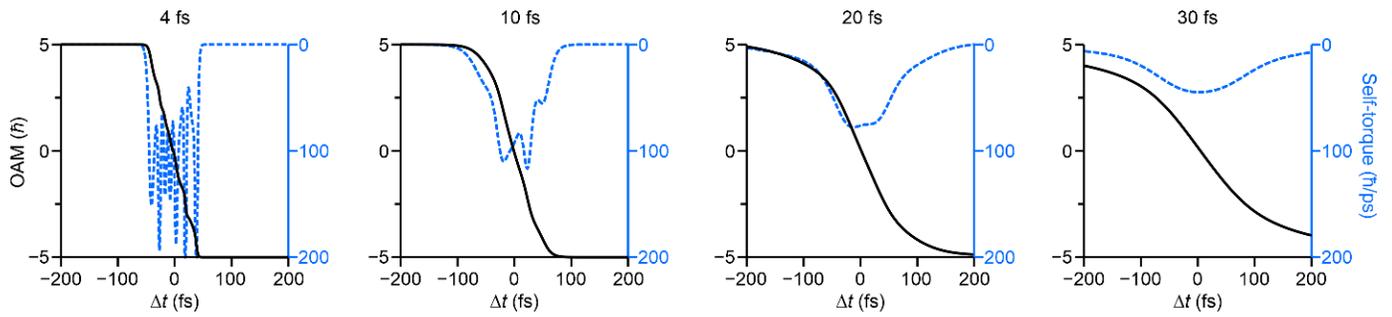

**Extended Data Fig. 3 | Dependence of orbital angular momentum and internal torque on the temporal coherence length.** Black curve, orbital angular momentum; blue dashed curve, internal torque; top labels, temporal coherence length [37]. If the coherence time is short, the different sidebands have well-defined arrival times, and the total orbital angular momentum evolves in steps, causing spikes for internal torque. For larger coherence times, the sidebands overlap in time and the internal torque becomes smooth.